**UNIVERSITY OF HERTFORDSHIRE**

**School of Computer Science**

**Modular Masters Programme in Computer Science**

**7COM1039 Advanced Computer Science Masters Project**

**Final Report**

**May 2015**

**A comparison of the performance and scalability of relational and document-based web-systems for large scale applications in a rehabilitation context.**

A. J. Williams, 11437073

## *Abstract*

**Background:** The Virtual Rehabilitation Environment (VRE) provides patients of long term neurological conditions with a platform to review their previous physiotherapy sessions, as well as see their goals and any treatments or exercises that their clinician has set for them to practice before their next session.

**Objective:** The initial application implemented 21 of the 27 core features using the Microsoft ASP.NET MVC stack. However, the two core, non-functional requirements were negated from the project due to lack of experience and strict time constraints. This project aimed to investigate whether the application would be more suited to a non-relational solution.

**Method:** The application was re-written using the MEAN stack (MongoDB, ExpressJS, AngularJS, NodeJS), an open source, fully JavaScript stack and then performance tests were carried out to compare the two applications. A scalability review was also conducted to assess the benefits and drawbacks of each technology in this aspect.

**Results:** The investigation proved that the non-relational solution was much more efficient and performed faster. However, the choice of database was only a small part of the increase in efficiency and it was an all-round better design that gave the new application its performance upper hand.

**Conclusion:** A proposal for a new application design is given that follows the microservice architecture used by companies such as Amazon and Netflix. The application is to be split up into four parts; database, client application, server application and content delivery network. These four, independently scalable and manageable services offer the greatest flexibility for future development at the low costs necessary for a start-up.

## *Acknowledgements*

I would like to take this opportunity to thank those who supported me throughout this project and my time at university: Dr Wei Ji and James Malcolm for supervising throughout this project, my parents and grandparents, Sharon and Christopher Williams and Norman and Ann Green for supporting me in everything that I do in life, my girlfriend, Grace Dark for always being able to re-motivate me during the longest of hours and finally to Dr Austen Rainer for his support throughout my final year of university, even after emigrating to the other side of the world. Thank you.

# Contents

# 1. Introduction and Overview

## 1.1 Summary of the research

In England, 110,000 people a year suffer a stroke (NHS [II], 2013), meaning that any application designed, that plans to aid the rehabilitation of strokes, and of other long term neurological conditions (LTNC), needs to be able to handle a large number of users. However, the risk of any application failing means that putting an infrastructure in place to handle this amount of traffic from the offset is not viable. Instead the application needs to be able to start off on a smaller scale and then be cost effectively able to grow naturally as the number of users of the application increases. This report gives a comparison of performance and scalability in relational and document-based databases, for large scale applications in a rehabilitation context.

## 1.2 Summary of the problem

The contact time between patients and clinicians drops significantly as a patient transposes from an inpatient to an outpatient, following the diagnosis of a long term neurological condition. To tackle the sense of abandonment the patient may experience during this transition, a health researcher at the University of Hertfordshire proposed a web application solution. The applications purpose is to serve as a platform to enable all parties involved in the care cycle to communicate, share information and allow patients to review their targets and exercises.

At present, the application contains the essential functionality needed to manage a patient, with the view of expanding and eventually serving as a foundation for an application that is the centre of all LTNC rehabilitation. This means integrating more features and technologies such as; being able to host video chats, instant exercise feedback using the Xbox Kinect sensor and video & audio editing, so that clinicians can customise how content is displayed to the patient. For example, setting a sequence of videos to play for a patient may involve playing the first ten seconds of a first video ten times and then playing a second video three times while a separate audio file plays over the top that explains the exercise.

The application was built by a team of six students using the C# .NET MVC stack with a SQL Server database. The project did not deliver all of the functionality it planned to and questions were raised as to whether the application met the desired performance and scalability requirements. Would it be more suited to a NoSQL solution?

The existing application will serve as one of the compared applications in this report, whilst the second will be a new application developed alongside the report. It will be developed using the MEAN stack, a full JavaScript stack consisting of MongoDB, ExpressJS, AngularJS and NodeJS. A performance and scalability review will be undertaken in an attempt to decide which application is best suited to its desired purpose.

## 1.3 Practical investigation and deliverables

There were 27 core features identified in the original project, with the existing application successfully implementing 21 of them. This project will be completed in less time than the first and there shall be only one developer working on the application. Therefore, the project only aims to implement 15 of the 27 core features. Even by only developing 15 features, it will still mean that the project is able to deliver features at a faster rate than the original application. This is mainly due to two reasons; the rapid development nature of the MEAN stack (explained in section 1.4) and as the project is a duplicate of an existing application, the requirements are well defined and unlikely to change throughout the project. However, this project will be paying less attention to the testing and non-core features such as the search functionality.

The application has separate interfaces for the three different types of user; patients, clinicians and administrators. Administrators manage all user accounts and a central repository of resources. Clinicians have a set of patients assigned to them to which they can allocate goals and treatments. Content, from either the central repository or directly from the clinician's machine (which can be uploaded) can be assigned to treatments to aid the patient in achieving their goals. A clinician can also assign content to an information section which gives general information on living with the particular condition. Patients will have a 'read only' view of their goals, treatments and information sections. They will also have the ability to comment on their goals to aid discussion on their progress with their care team.

This project will mainly be developing and testing the clinician interface as it presents the most transactional logic and has a wide variety of features. There are 11 features in the clinician interface of which the project will be implementing 9. In addition to this, all four administrator features will be developed as they are necessary for the clinician interface. The 6 features that make up the patient interface will not be implemented. A summary of the features is given in Appendix A.

## 1.4 Tools and techniques

The MEAN stack is a full JavaScript framework and acronym for MongoDb, ExpressJS, AngularJS and NodeJS (Karpov, 2013).

**MongoDB** - an open source document database that was designed to be highly scalable. (MongoDB, 2015).

**ExpressJS** - a lightweight MVC framework that sits on top of NodeJS (ExpressJS, 2015).

**AngularJS -** a front end JavaScript framework for quickly developing single page, dynamic web application (AngularJS, 2015).

**NodeJS** - a event-driven I/O server-side JavaScript environment built on Chrome's V8 JavaScript runtime for building fast, scalable network applications (NodeJS, 2015).

The image below shows how the components work together to make up an application:

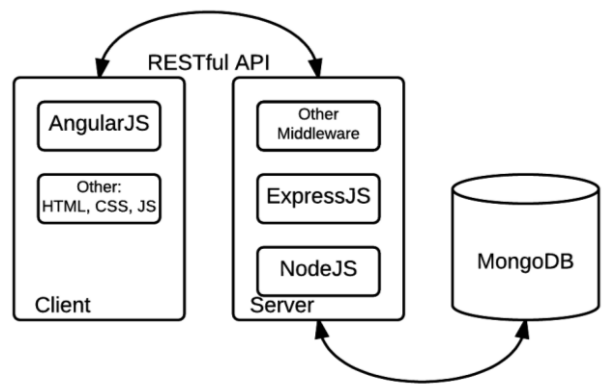

NodeJS is a core framework that includes only the base functionality for the server. Developers then include modules into their Node application, in order to tailor the application to its specific purpose. For example, the ExpressJS module gives a lightweight MVC platform for web applications, the Mongoose module (Mongoose, 2015) provides the entire object modelling functionality for interacting with the database and PassportJS (Hanson, 2015) provides all the authentication and account functionality. Although not quite conforming to the traditional definition of the word, the MEAN community refers to these third party modules as middleware.

The whole application is written in JavaScript, this therefore allows developers to create robust applications quickly as knowledge of only one language is needed and developers do not need to switch between languages. The MEAN stack promotes API first development which suits the application; this is because the eventual intention is to allow new technologies to plug into it. Another benefit is that API first development follows the microservice architecture, a method that is ideal when an application needs to support an array of different devices (Lewis and Fowler, 2014). This works through the use of a server side application with a public API sitting in the cloud. This application is called a microservice. Multiple client applications can then be developed that use this API and any others if desired. Therefore, this means that the core functionality of the application does not have to be re-written for each device (Desktop, Phone, Watch, etc.). Designing the application in this way means that not only is it ready for future scaling, but also other rehabilitation tools can use the VRE at an API level. An example of a future use could be a rehabilitation games device that uploads the results to the VRE so that clinicians can monitor progress.

## 1.5 Research questions

The overall objective of this project is to compare the performance and scalability of applications using relational and non-relational databases. Furthermore, a general evaluation of the particular applications needs to take place, in order to give the comparative context. To achieve this, the objective is refined into three research questions, with question two being further broken down into two sub questions:

- RQ1: What are the particular technical constraints and requirements of the type of problem described?
- RQ2: What is the relative impact of relational and document database design on overall system performance?
  - RQ2.1: What performance issues are there with the document-based database?
  - RQ2.2: What performance issues are there with the non-database aspects of the system
- RQ3: What is the relative impact of relational and document database design on overall system scalability?

RQ1 and RQ3 will be investigated using qualitative, active research techniques, giving an overall review to answer the questions (Petersen et. al, 2014). RQ2 will be investigated by developing the new application and then using NeoLoad (Neotys, 2015) to run performance tests against the two applications. The results will then be analysed and an evaluation will be given.

## 1.6 Motivation for the project

The project brings an opportunity to investigate whether the application would be better suited to a NoSQL solution at this early stage of the product's development, before the existing application is released to clinicians. Continued involvement in the existing application means that an alternative solution can be offered, which can be implemented with a better understanding of the requirements. Therefore, the application should be more carefully thought out, hence better designed.

On a more personal level, the application allows the researcher to deepen their knowledge of these technologies which are widely sought after in the current web development industry.

## 1.7 Report structure

The report will be structured as follows:

1. **Introduction** – an introduction to the report and project. The research questions are stated as well the motivation for the project.
2. **Description of the problem** – a detailed explanation of the application developed and the rehabilitation project.
3. **Previous work review** – a literature and technology review of previous work and techniques.
4. **Overview of the proposed solutions** – a technical description of the existing and new solutions to explain the differences and similarities in their design and architecture.
5. **Investigation** – the research questions proposed in section 1.5 are investigated and answered.
6. **Discussion, evaluation and conclusion** – a collation of results and summary of the project in its entirety.

# 2. Description of the problem

## 2.1 Problem domain

It is not uncommon for contact time between a patient and their clinical care team to drop significantly when transitioning from an inpatient to an outpatient. Sufferers of stroke for example, can go from having six hours a day of physiotherapy as an inpatient to less than one hour per month as an outpatient. This leads to the patient feeling abandoned and frustrated that their rehabilitation is not as quick as it possibly could be. A solution proposed to tackle this issue is the application which this project uses to investigate and answer the questions stated in chapter one. It has been named the Virtual Rehabilitation Environment (VRE).

## 2.2 Solution functional description

The VRE is a web application with an aim to provide an environment much like the virtual learning environments that many universities provide for their students, but with a focus on rehabilitation and the particular requirements needed for sufferers of long term neurological and physical conditions. Eventually the goal is to be able to use the VRE as an active tool for providing patients with games and exercises to aid their rehabilitation and provide instant feedback while tracking overall progress. However, while progress is being made by researchers worldwide in using the latest technologies to aid rehabilitation, the healthcare industry appears to have missed the step that comes before that, which is to have a central information system to run these developments from. The VRE planned to address this first by initially creating a foundation application. Each patient has a profile to which there are three main sections:

**Goals** – A patient can have many goals to work towards over varying durations. All stakeholders of the patients' care including the clinical team, carers and family meet periodically to review and set these goals. Then all care, physiotherapy and treatments are used to help achieve these targets.

**Treatments** – Treatments are set by clinicians and are usually exercises to be practiced a set amount of times per day. Currently, exercises are set during contact time and aided by an array of simple materials supplied by the clinician such as drawings of stick men. The VRE uses a central repository of resources such as video exercises, images and audio descriptions. Content can be assigned to a treatment from the central repository to aid the patient in practicing in-between sessions. If no appropriate content exists then the clinician may upload their own content to a patient's treatment. A big advantage of this is that clinicians can record entire sessions with the patient which can then be played back and practiced by the patient at their own disposal.

**Information** – The information section serves as a place for clinicians to give more general advice to their patients. This could be a generic information pack on the patient's condition or a guidance booklet on how to aid someone in a wheelchair in using a toilet.

Any clinician associated to a patient can upload and make changes to the patient's profile. The patient has a read-only view of the goals, treatments and information assigned to them and can review this material, allowing them to work more effectively and less from the memory of what was said to them in their last session.

Neurological conditions can affect each patients abilities differently based on a number of factors; type of condition, severity, area of the brain affected etc. For this reason it is important that clinicians can tailor each patients interface to their own particular requirements. A patient struggling with motor functionality may need larger buttons on their screen whereas someone suffering from visual field loss (Windsor, 2015) may have complete motor control but be only able to see half of the screen. Although the VRE gives clinicians a central location to manage all rehabilitation patients, it is essentially adding to the clinician's workload. For this reason, a main requirement is that it is incredibly simple and natural to use. For the VRE to be of use to clinicians, it needed to seamlessly fit in with their day to day workflow.

## 2.3 Application users

The VRE currently has three different user roles, each with their own interface:

**Administrators** – VRE system administrators are responsible for the maintenance and structure of the central repository. They also manage all accounts and are currently the only user group who can create other users. In the future it is expected that a tiered administration would lead to district administrators being able to create patient and clinician accounts for their local trusts, whilst the central repository is still managed by a core team of VRE staff.

**Clinicians** – Clinicians are assigned to patients and each clinician is presented with a list of their patients upon log in. One patient may have many clinicians assigned to them. For each individual patient, the clinician may manage their goals, treatments and information. Treatments and information can be accompanied by content from the central repository or uploaded directly from the clinicians' computer. The clinician is also able to browse the repository and search for content.

**Patients** – Patients are presented with an interface that has been configured to their individual needs by their clinical team. On logging in, they are presented with a welcome screen that displays basic information such as their name and the time. Patients can view their goals, treatments and information and comment on how they are progressing to communicate with their clinician. Content assigned to treatments and information by their clinical team can also be viewed.

## 2.4 Core stakeholder

Ms Sally Davenport, a health researcher at the University of Hertfordshire is the person who initially proposed the VRE project. Ms Davenport saw a gap in the industry for a base information system to aid patient rehabilitation and after working on the concept with a computer science PhD student, asked for an application to be developed quickly so that she could start pitching the project to local clinicians to get feedback. Ms Davenport has worked with the developers in a product consultant role to ensure the requirements she envisaged have been met.

## 2.5 Scope of this project

The application has already been developed as part of an earlier project and delivered to Ms Davenport. However, a discussion on the technology most appropriate for the application was not undertaken during the initial project and instead the technology that the most students were proficient in was used. Proficiency is a major factor in ensuring an application is delivered on time and to a high standard but there are other factors in the VRE project that were negated from the initial project, namely performance and scalability. An application that aims to serve hundreds of thousands of people each year must be able to handle the large amount of traffic and be able to scale easily. This project will re-develop the application using the MEAN stack, a technology stack deemed more scalable and efficient by the web community. The project will then test the performance and review the scalability. This will allow the researcher to make an informed decision on whether the MEAN stack is more suited to the VRE application than the current ASP.NET solution.

# 3. Literature and technology review

The relational model proposed by Edgar Codd in 1970 (Codd, 1970) continues to dominate nearly all businesses and organisations to this day. However, the rapidly growing amount of data that these companies now wish to store leads to difficulties when it is stored in a relational database. Horizontally scaling the database is problematic and querying the data becomes slow and inefficient. This has led to a recent rise in popularity of non-relational databases. Non-relational or 'NoSQL' databases boast their ability to be flexible, horizontally scale easily and maintain high levels of performance by sacrificing the ACID (Barry, 2015) properties of relational databases for what is called BASE properties (Basic Availability, Soft state, Eventually consistent). This is not ideal for all solutions and so sometimes a relational model is necessary but most applications can afford to relax the strict rules enforced by relational databases. Non-relational databases predate Codd's paper but since becoming popular, much research has been done in comparing the abilities of the relational and non-relational model.

Yahoo researchers developed the Yahoo Cloud Services Benchmark (YCSB), a framework for measuring the performance and scalability of database management systems (Cooper et. al, 2010). This benchmark has since been used in many studies to evaluate the performance of multiple technologies such as: (Abramova et. al, 2014), (Tudorica and Bucur, 2015) and (Abubakar et. al, 2014). An independent study by United Software Associates found that MongoDB provides better performance than its main competitors; Couchbase (Couchbase, 2015) and Cassandra (Apache, 2015), achieving better results in all tests using the YCSB by as much as 13x (USA, 2015). However, Mongo has been criticised about how it achieves these benchmarks. For example, when a write is made to Mongo, it is staged and waits to be written to the database to be stored permanently (MongoDB [III], 2015). The problem here is if there is an issue with the writing of the data, the data is lost without notifying the user.

There is little published work that directly compares MongoDB with SQL server but a blog post by Michael Kennedy (Kennedy, 2010) shows that on having 5 concurrent operations inserting 10,000 records each, for one of these operations the SQL database took 204.215 seconds whereas Mongo only took 2.032. Also, when querying the data back, the SQL database took 28 seconds to query 50,000 records whereas Mongo took only 10.4 seconds. In all tests, Mongo performed significantly better. However, a 2013 master's thesis from a student in the University of Edinburgh comparing Mongo and MySQL (Hadjigeorgiou, 2013) identified that although Mongo out performs the RDBMS on inserting data, when deleting records, it was MySQL that completed its operations quicker. This thesis found that on average, Mongo performed around 40% better then MySQL and is able to execute a larger number of queries in a smaller amount of time. In contradiction to both of these experiments, an article in the 2012 international conference of very large databases found that this wasn't the case. A study was done against Mongo and SQL server using the YCSB and the TPC benchmark (TPC, 2015). The results found that SQL server was still able to outperform Mongo, although they state that Mongo's performance has continued to increase over time (Floratou et. al, 2012). Two of the five working on this study were Microsoft researchers. Therefore, it is possible that their opinions were bias due to the personal involvement with the product in question influenced results. Also, being more familiar with SQL server, they may have applied a higher level of optimisation than their Mongo implementation.

It is understood that NoSQL databases perform better than databases that follow the relational model but in return sacrifices are made on the ACID properties of the data. However, this has not prevented over a third of the Fortune 100 from using MongoDB in their organisations (MongoDB [IV], 2015). Expedia uses Mongo to provide users with its ScratchPad tool, an application that allows users to plan their travel and save searches across multiple devices. Mongo's flexibility and rapid development allowed them to prototype the application in just two months and put the application in production within five months. During this time, they had to radically change their schema three times (MongoDB [V], 2015). Similarly, Weather Channel upgraded all their legacy systems to use Mongo. This allows them to iteratively implement features in just hours instead of the weeks it would take before the change (MongoDB [IV], 2015). However, in some organisations where it would seem to make sense to use NoSQL given their vast amount of data, still use RDBMS's for their applications. CERN use Oracle Exadata servers for its BigData and Oracle Real Application Clusters to manage raw experimental data stored in ROOT structured files (Heath, 2012). EBay also use Oracle but do not enforce referential integrity as it would drain the performance too much. Instead the application code keeps the data consistent and maintains all foreign keys (Shoup and Pritchett, 2006). Other companies that manage large amounts of data such as Google and Amazon have written their own database management systems (Googles BigTable and Amazons Dynamo) to tailor their databases to their own unique requirements (DeCandia et. al, 2007). Facebook and Twitter use a mix of NoSQL and relational databases but both use MySQL as their primary database, with Facebook boasting the one of the largest MySQL clusters in the world (Priymak, 2013).

Many large organisations such as Amazon have their primary application made up of multiple small services each providing a subset of the applications functionality (Rohde, 2007). This 'microservices' architecture sees a larger application made up of multiple, independently scalable services that interact with each other using API's (Lewis and Fowler, 2014). Today, even smaller websites are making use of the performance gains this brings by using Content Delivery Networks (CDN) to provide stylesheets and JavaScript libraries to their applications. CDN's replicate content over a distributed network but give it only one logical address (Sexton, 2015). This means that users accessing the website can pull the content from the server that is closest to them, shortening the time taken to process the request. Google use this method to deliver YouTube videos to users (Torres et. al, 2011).

To conclude, a lot of organisations today are willing to sacrifice the consistency of a relational model for a more flexible data structure, as it allows them to develop features quicker. As the size of web applications and the amount of data stored grows, the performance and scalability benefits of a non-relational application out-weigh the need for a strict data structure the majority of the time. This isn't the case in critical systems such as banking applications and patients medical records where the data has to be always consistent and up to date. In these instances, if the application is large enough, organisations use a mix of databases so that their critical data can be consistent and their non-critical data can take advantage of the NoSQL benefits.

# 4. Overview of proposed solutions

## 4.1 Existing relational solution

The original application was developed using ASP.NET's C# MVC3 pattern. This takes a bottom-up approach, designing the database first and then building the application on top to interact with it (FitzMacken, 2014). The database, stored on a Microsoft SQL Server, is first designed and created. Then the models are created automatically in Visual Studio by analysing the tables using Entity Framework (ASP.NET [II], 2015). The models may be changed manually to add validation but VisualStudio is good at maintaining referential integrity and any database constraints that are in place. Unlike non-relational databases where collections are made as they are needed, relational databases have to be planned and designed carefully. A lot of planning was carried out to ensure the database would be ready for future development to be able to take place without needing to make any significant changes to the database. For example, the content was given an inheritance structure. Video, Audio and Text tables inherit common characteristics from the Content table. As the Text and Audio content had no specific metadata, these tables just contained the ID of the parent content, but were designed this way in case future development required specific meta data to be added.

There is little client-server separation; instead the Razor view engine (FitzMacken, 2014) renders the view on the server side before pushing out to the client. This means that unlike traditional web development, logic can be put into the views and reduce the amount of requests made to the server. The controllers sit between the views and the models, housing the majority of the applications logic. The controllers provide functions to the view that interact with the models to make changes to the database.

This application is made up of 9,820 user generated lines of code and 580,103 lines with all the middleware included. However, it is difficult to make a direct comparison between the two applications on just the lines of code for two reasons. Firstly, the existing application implements more features of the core application than the new application. A full comparison and feature catalogue can be found in appendix A. Secondly, the existing application uses very few libraries, much of its functionality is unnecessarily reinventing the wheel by re-implementing code better implemented elsewhere. The new application aimed to use as many existing libraries as possible so that it could be developed as quickly as possible.

A more in depth look at the existing system design and technologies is given in the final report submitted for the team development exercise module that the application was developed for. Appendix B gives a URL for this report.

## 4.2 New non-relational solution

The new application is developed using the MEAN stack, a complete open source stack written in JavaScript. The name MEAN is an acronym of the stacks components, MongoDB, ExpressJS, AngularJS and NodeJS.

### MongoDB

Mongo is an open source, non-relational document database that has become very popular over recent years. Mongo boasts the ability to scale up an application quickly and cost effectively by being able to just add more severs. Unlike relational databases, the data is stored as collections in key-value pairs reminiscent of associative arrays. This simpler and more flexible method for storing data allows for quicker development and allows schemas to evolve effortlessly (MongoDB, 2015).

**NodeJS**

NodeJS is a platform written on Google Chromes V8 runtime engine (NodeJS, 2015) that allows for server side JavaScript. Node is asynchronous, giving its concurrency through non-blocking code, asynchronous callbacks and an event loop. Phillip Roberts gives a detailed explanation of this in his 2014 JSConference talk (Roberts, 2014).

**ExpressJS**

Node only comes with its core library as standard and requires third party libraries to be installed on top of it for most functionality. Express is a lightweight MVC library that is installed on top of Node (ExpressJS, 2015). Other libraries are used in the application to provide extra functionality such as Mongoose for object modelling (Mongoose, 2015) and PassportJS for the VRE's authorisation and authentication (Hanson, 2015).

**AngularJS**

Angular is a framework developed by Google that aids the development of complex, single-page interfaces and enables fast development (AngularJS, 2015). Custom directives allow developers to create their own HTML elements and attributes, which reduce duplication and allow for complex interfaces to be understood quickly. For example:

```html
<modal title="Add Category" visible="addCatModal">
    <div addcat></div>
</modal>

<modal title="Edit Category" visible="editCatModal">
    <div editcat></div>
</modal>

<modal title="Add Content" visible="addConModal">
    <div addcon></div>
</modal>
```

Here, a custom 'modal' element has been created that means the code does not have to be repeated three times. Inside each modal, a custom attribute has been created that pulls in the content for each particular modal. A developer looking at this can see clearly what it is that the code is doing. Traditionally, HTML bookmarks are used to navigate users to a specific part of a large webpage. For example, *http://getbootstrap.com/components/#nav* directs you to the navigation section on the Bootstrap components page. Angular takes the rest of the URL after the hash symbol and parses it so that it can simulate routing. For example, *http://localhost:3333/#!/Administrator/repository* navigates the user to the repository management page on the administrator interface. This allows for complex single page application to still benefit from the separation and structure of traditional multi-paged web applications (AngularJS [II], 2015).

The MEAN stack promotes good design by forcing the developer to completely separate the client and server side components using a RESTful API. What this means is that client applications communicate with their server side counterparts using the HTTP request methods; GET, POST, PUT and DELETE (Fielding et. al, 1999). Decoupling an applications server and client side development encourages API first development. This therefore benefits today's multi-platform web as it forces developers into carefully designing their applications, enforcing separation of concerns and hiding internal complexity (API First, 2013). Designing the application this way prepares the application for future scaling by naturally following the microservice architectural style. Microservice architecture is a relatively new term in the web development community, used to describe a particular way of designing software applications as suites of independently deployable services (Lewis and Fowler, 2014). Using this architecture means that not only is the application made up of multiple services that can scale independently of the others, but also that multiple interfaces can be written to use the same server side API's. This means that complex functionality at the applications core does not need to be re-written for each device the application is to be deployed on. For example, writing a new interface for the VRE to run on a smart watch is now just a case of writing the user interface and plugging it into the existing API.

NodeJS has a native '*route*' function to register the applications routing and create the REST API. The routing has been split out into multiple JavaScript pages, each handling one controller and one collection in the database. For example, here is the page that handles the goals management routing:

```javascript
var goals = require('../../app/controllers/goals.server.controller'),
    accounts = require('../../app/controllers/accounts.server.controller');

module.exports = function(app){
    app.route('/api/goal')
        .get(goals.list)
        .post(accounts.requiresLogin, goals.create);

    app.route('/api/goal/:goalId')
        .get(goals.read)
        .put(accounts.requiresLogin, goals.update)
        .delete(accounts.requiresLogin, goals.delete);

    app.param('goalId', goals.goalById);
}
```

At the top, the goals and accounts controllers are pulled in as we want to be able to call functions from these controllers when the user navigates to a particular route. If a user sends a GET request to '*/api/goal'* then the '*list'* function in the goals controller is called and a list of all goals returned. If a user sends a GET request to '*/api/goals/55ad34ea83922e001d3c50bd'* then the '*read'* function in the goals controller is called and the goal with the particular ID stated in the URL is returned. The other request methods all require the '*requiresLogin'* function to return true before calling a function in the goals controller. This is a security feature that prevents anyone from being able to make changes to the application. The GET requests do not require a user to be logged in for testing purposes but once in production, with the application dealing with patient data, all of the API will require users to be logged in.

The AngularJS front end is created using a hierarchy of modules, this design allows for functionality to be made generic to the entire application or specific to a particular module by just changing where it is declared. At the top of the hierarchy is the application that holds the entire applications configuration. Underneath that is a 'Main' module that holds functionality generic to the entire application, such as the modal directive that handles the applications modal windows. The next step down the hierarchy is where the three interfaces are declared: administrator, clinicians and patient. To interact with the REST API, the application uses Angulars built in factory pattern and ngResource (AngularJS [III], 2015). NgResource is a tool built on top of Nodes $http service to easily interact with REST back ends. Below, the $resource service is passed into the function and used to interact with the API. The $resource

service comes with five methods as default: *get*, *query*, *save*, *remove* and *delete*. To be able to use the *update()* method, it is registered manually.

```javascript
angular.module('clinician').factory('Goals', ['$resource', function($resource) {

    return $resource('api/goal/:goalId', {
        goalId: '@_id'
    }, {
        update: {
            method: 'PUT'
        }
    });
}]);
```

Now, in the controllers the 'Goals' service is declared and used like so:

```javascript
$scope.updateGoal = function(){
    var go1 = new Goals({
        description: $scope.goal.description,
        term: $scope.goal.term
    });
    Goals.update({goalId:$scope.tmpGoalId}, go1, function(){
        //alert("update ok");
        $scope.success = "Goal updated";
        $scope.toggleEditGoalModal();
    }, function(errorResponse){
        $scope.error = errorResponse.data.message;
    });
}
```

Here a new goal is declared using the data sent from a user submitted form and then the *update()* method is called on the Goals service. This method takes in: the ID of the goal to be updated, the newly declared goal and then the functions for the methods success and failure. Appendix C gives a complete walkthrough of the repository upload process. This shows how all the components of the MEAN stack interact together.

The application uses a separate model, controller and routing page for each collection in the database. The client side then has a service to interact with each routing page. The collections stored in the database are:

- Accounts
- Administrators
- Categories
- Clinicians
- CliniciansPatients
- Contents
- Goals
- Information
- Patients
- TreatmentContent
- Treatments

The application contains 5,696 lines of code, 1,861 of these are server side and 3,835 make up the client application.

# 5. Investigation

## 5.1 RQ1: What are the particular technical constraints and requirements of the type of problem described?

Non-functionally, it was a definite that the application had to be able to perform well under a large traffic load on the server and that it should be able to scale naturally as the number of users increases. Other non-functional requirements were that it should be secure as it is storing patient data and that it provided clinicians with a naturally easy to use interface because it had to fit in with their busy schedule. The last of these requirements ended up being the easiest to achieve. During the initial project, the team were under a tight schedule and pressure to demonstrate new functionality to Ms Davenport every week. To develop the user interface quickly, they chose to use Twitter's Bootstrap framework. Instantly, Ms Davenport said she was happy with the applications appearance and when demonstrated to clinicians in focus groups throughout the project, they too praised its simplistic easy to navigate design. For these reasons, the default Bootstrap style became the standard for the application. Some of the more difficult functional areas of the relational application such as the central repository were styled from scratch but the new, non-relational solution solely uses Bootstrap throughout.

To ensure a high level of security in the application, minimum personal data is stored on each patient to comply with the NHS information governance (NHS, 2015). Both applications store randomly created salts to hash the passwords. The relational application uses the SHA-256 algorithm. This stands for Secure Hash Algorithm and is "a hash algorithm used by certification authorities to sign certificates and CRL (Certificate Revocation Lists)." (TBS Internet, 2015). This algorithm generates a hash value consisting of 256 bits. A random salt is used to ensure each hash is different, even if two passwords are the same. The relational application stores the salt in the same field as the password in the database like so:

$$(unhashed)salt + (hashed)(salt + password)$$

The non-relational application uses the local strategy provided by PassportJS (Hanson, 2015) for its authentication. It uses Node's 'Crypto' package (NodeJS [II], 2015) to hash the passwords using its 'pbkdf2syn' function. Standing for 'Synchronous, Password-Based Key Derivation Function 2,' this function applies a pseudorandom function such as a cipher to the password along with a salt and repeats the process a set amount of times to produce a key:

```
AccountSchema.methods.hashPassword = function(password){
        return crypto.pbkdf2Sync(password, this.salt, 10000, 64).toString('base64');
};
```

In the application, the process is iterated 10,000 times to find the final hash. By default this function uses SHA1, but setting this to use SHA256 is just a case of adding it as a parameter in the above function.

The final two non-functional requirements are performance and scalability. The motivation for this project was to find out if the application is more suited to a different technology to the existing ASP.NET stack and that is investigated in RQ2 and RQ3. However, re-writing the application gave opportunity for other changes that contribute to these factors. The repository for the initial application is a hardcoded directory on the C drive of the server that the application resides on. This means that once the application is compiled, it cannot be changed. Also, the application server would very quickly reach its full storage space capacity and need to be scaled out. For the new application, it was designed to work with a content delivery network (CDN). Content delivery networks are a set of distributed servers that provide content to clients based on the geographic location of the user (Zakas, 2011). For example, a person trying to view a video in the south of England retrieves it from a server in the London datacentre where as a person trying to view the same video using the same URL in Scotland is given the video from the Glasgow datacentre. Replicated content is managed in a similar way to database replication servers. Once a change is made, it is pushed out to all other copies of the content. This benefits the application from a performance perspective as there are several copies of the content allowing clients to retrieve from a geographically close server, rather than one that is located at the other side of the country. Also, this is good for scalability as the CDN has the ability to scale independently of the rest of the application. The way that the new application is designed is that a base URL is provided in a configuration file, stored on the server:

```
module.exports = {
    //Development configuration options
    db: 'mongodb://localhost/vre-dev',
    sessionSecret: 'developmentSessionSecret',
    VRE_GLOBAL_REPOSITORY: 'C:\\Users\\ash\\Desktop\\VRE\\public\\VRE_REPOS'
};
```

For development purposes, the VRE_GLOBAL_REPOSITORY variable is set to an area on the C drive, but this can easily be changed to the URL of the CDN (e.g. cdn.vre.com). Then, each item in the repository stores its full path based on this variable:

| path | cat |
|---|---|
| C:\Users\ash\Desktop\VRE\public\VRE_REPOS\0E5OBOH0z307eR2Eocg5APMb.mp3 | 55 |
| C:\Users\ash\Desktop\VRE\public\VRE_REPOS\44Db_7yMLy51_FvPKBe47g-1.mp4 | 55 |
| C:\Users\ash\Desktop\VRE\public\VRE_REPOS\CBiI3I8krVRPAHaGRC0tCTlr.pdf | 55 |
| C:\Users\ash\Desktop\VRE\public\VRE_REPOS\wC5RPOF59dRJUGKT8Wo5DAUh.docx | 55 |
| C:\Users\ash\Desktop\VRE\public\VRE_REPOS\ZGy5H2bgT4n7Q6W-oUhgUaT8.JPG | 55 |

Another scalability requirement from the initial application was it needed to be easy for future developers to build on top of it. In the 2014 F8 conference, Tom Occhino, an engineering manager at Facebook talked about how MVC doesn't scale as it becomes very complicated, very quickly. This led to new engineers coming into teams finding it difficult to get up to speed with the application, and being terrified that if they made a change, it would break something that they didn't expect (Occhino, 2014). This talk was met with a backlash from the web developer community who blamed Facebook for using the MVC pattern wrongly. However, there were similarities appearing with the existing application. The entire applications functionality was written in 6 controllers and the views were rendered on the server before being sent to the client so there was no client-server separation. The new application has split the server and client application totally; they interact with each other through an API provided by the server. Both the server and client application have their own controllers (currently 13 each). Splitting the application down like this makes it clear where functionality should be physically located and therefore easier for developers to make changes to the code.

Technical constraints in both projects mainly revolved around time and money. The hardware used to test the performance of both applications was a consumer laptop with an Intel i5 processor and 8GB of memory. Although sufficient enough to test each solution for differences, the tests are not a true reflection of how the applications would perform in a production environment. Furthermore, time and resource constraints have meant that testing the scaling of each solution has not been feasible. Instead, RQ3 has been investigated using a diagnostic active research technique described by Kai Peterson et al (Peterson et al, 2014). This technique requires the researcher to first understand the problem and then take action based on that understanding. This project has been unable to take action so instead a review the options will be conducted and a solution proposed, conforming to stages 1 and 2 of Peterson et al action research cycle.

## 5.2 RQ2: What is the relative impact of relational and document database design on overall system performance?

### RQ2.1: What performance issues are there with the document-based database?

To investigate this research question, a performance test was carried out on both the existing application with a SQL Server backend and the new application with a MongoDB back end. The investigation is described below.

**Environment:**

Each application was set up on the same test machine.

Test workstation & server:
- Windows 8.1 Home edition
- Intel i5 Processor
- 8GB RAM

Existing application:
- SQL Server 2012
- Internet Information Services version 6.2

New application:
- MongoDB version 3.0.3 2008 R2
- NodeJS version 0.12.3

Test Software:
- NeoLoad version 5.1.0

All of the technology is installed on the same workstation as where the tests were run to ensure both environments are as similar as possible, and to eliminate the risk of network traffic interfering with the results. Both databases were standalone, non-distributed solutions. See section 5.3 for a scalability review of each technology.

**Method:**

NeoLoad will be used to set up an array of virtual users that can interact with the application, each virtual user will be able to perform one pre-recorded, functional operation in the application. Scenarios will then be set up to simulate multiple concurrent users performing their functional operation. NeoLoad includes features to add conditional logic and looping in the virtual user scripts. An example of a test may be to have 25 concurrent users, add 10 patients each and then assigning another clinician and adding a new goal. Another example could be to simulate 50 concurrent administrator users trying to create 500 clinicians each.

### *Virtual Users*

These are simulated users created by using NeoLoad to record interactions with the application. Each user is a recorded interaction with a certain part of the applications functionality and then further logic can be added to loop actions and handle errors and conditional logic.

| Name | User | Description |
|------|------|-------------|
| Add100 | Administrator | Log on as administrator, add 100 clinicians, log out |
| Goal100 | Clinician | Log on as clinician, view patient, add 100 goals, log out |
| View100 | Clinician | Log on as clinicians, view 100 patients, log out |
| UpdateRep100 | Administrator | Log on as administrator, update the name of 100 repository items, log out |

### *Scenarios*

'Scenario' is the term given to a test run in NeoLoad. The user creates a population from the list of recorded virtual users and uses this population to run scenarios, specifying the number of concurrent users and number of iterations that the test should run for.

| ID | Virtual User | Concurrent Users | Description |
|----|--------------|------------------|-------------|
| 1 | Add100 | 10 | 10 concurrent runs of the Add100 user, leading to 1000 new clinicians added |
| 2 | Goal100 | 10 | 10 concurrent runs of the Goal100 user, leading to one patient having 1000 new goals |
| 3 | Read90/10 | 10 | 10 concurrent runs, 90% running the View100 user, 10% running the Goal100 user |
| 4 | Update100 | 10 | 10 concurrent runs of the UpdateRep100 user |
| 5 | Write50/50 | 50 | 50 concurrent runs, 50% running the Goal100 user, 50% running the View100 user |

### Results:

The following results tables give the statistic name, the result for the relational solution, the result for the non-relational solution and then the percentage difference between the two results. The results are discussed and evaluated in chapter 6.

### *Scenario 1*

Scenario one saw 10 concurrent users, each are adding 100 clinicians to the database. The non-relational solution took 51 minutes and 17 seconds whereas the relational model took only 2 minutes and 43 seconds.

| | SQL | MEAN | % |
|---|-----|------|---|
| Average pages/s | 12.4 | 5.3 | -57.3% |
| Average requests/s | 19.0 | 27.0 | +42.1% |
| Total pages | 2040 | 16220 | +695% |
| Total requests | 3110 | 82980 | +2,568% |
| Average Request response time | 0.456 s | 0.605 s | +32.7% |
| Total request errors | 0 | 0 | +0% |
| Error rate | 0 | 0 | +0% |
| Average Page response time | 0.691 s | 1.89 s | +174% |
| Total throughput | 13.52 MB | 99.89 MB | +639% |
| Average throughput | 0.66 Mb/s | 0.26 Mb/s | -60.6% |
| Total users launched | 10 | 10 | +0% |
| Total iterations completed | 10 | 10 | +0% |
| Total action errors | 0 | 0 | +0% |
| Alerts total duration | 0 % | 0 % | +0% |

The main thing to notice here is the 'Total Requests,' the relational application has only 3110 whereas the non-relational has 82,980. This is because the functionality to create an account was developed early on in the project when the researcher was still becoming familiar with the technologies. Instead of using Angular to display the new account, the page is refreshed to pull the new account from the database. The MEAN stack is for developing single page applications. This means that the majority of the application should be loaded into the client's browser when they first log in. In this case, refreshing the page means the client requests all of these pages again and therefore slowing down the execution dramatically. For these reasons, a similar test will be carried out for scenario 2.

### *Scenario 2*
Scenario two was chosen as the goals functionality was developed further into the project with a better understanding of how Angular works. Therefore, the same mistake made in the functionality tested in scenario one should not occur. Also, adding a goal is a simple form that creates a basic insert into a single collection. Its simplicity makes it the closest functionality the application gets to directly interacting with the database, as there is not much code between the user form and the insert into the database. Scenario two simulates ten concurrent users, each adding 100 goals to a single patient. The relational solution took 8 minutes and 32 seconds. The non-relational solution took only 32 seconds.

| | SQL | MEAN | % |
|---|---|---|---|
| Average pages/s | 6.0 | 39.1 | +552% |
| Average requests/s | 9.7 | 93.0 | +859% |
| Total pages | 3070 | 1290 | -58% |
| Total requests | 4956 | 3070 | -38.1% |
| Average Request response time | 1.03 s | 0.15 s | -85.4% |
| Total request errors | 226 | 0 | -100% |
| Error rate | 4.6 | 0 | -100% |
| Average Page response time | 1.65 s | 0.181 s | -89% |
| Total throughput | 422.06 MB | 1.07 MB | -99.7% |
| Average throughput | 6.58 Mb/s | 0.26 Mb/s | -96% |
| Total users launched | 10 | 10 | +0% |
| Total iterations completed | 10 | 10 | +0% |
| Total action errors | 0 | 0 | +0% |
| Alerts total duration | 5.5 % | 0 % | -100% |

### *Scenario 3*
Scenario three is designed to be read heavy with 90% read operations and 10% insert operations. For this, a NeoLoad population was created with the View100 user making up 90% of the traffic and the Goal100 user making up the extra 10%. The relational solution took 43 minutes and 41 seconds. The non-relational solution took 22 minutes and 56 seconds. Like scenario 1, this scenario required a page refresh each time a clinician views a patient and then clicks back to their patient list.

| | SQL | MEAN | % |
|---|---|---|---|
| **Average pages/s** | 1.2 | 11.3 | +842% |
| **Average requests/s** | 2.6 | 18.4 | +608% |
| **Total pages** | 3034 | 15555 | +413% |
| **Total requests** | 6860 | 25354 | +270% |
| **Average Request response time** | 0.58 s | 0.498 s | -14.1% |
| **Total request errors** | 11 | 0 | -100% |
| **Error rate** | 0.2 | 0 | -100% |
| **Average Page response time** | 1.31 s | 0.797 s | -39.2% |
| **Total throughput** | 899.8 MB | 19.98 MB | -97.8% |
| **Average throughput** | 2.75 Mb/s | 0.12 Mb/s | -95.6% |
| **Total users launched** | 10 | 10 | +0% |
| **Total iterations completed** | 10 | 10 | +0% |
| **Total action errors** | 0 | 0 | +0% |
| **Alerts total duration** | 0 % | 0 % | +0% |

### *Scenario 4*

Scenario four saw the names of 1000 repository items being updated using 10 concurrent runs of the Update100 user. Unfortunately, every time the test was run on the relational solution, the CPU usage hit 100% and crashed the workstation. The non-relational solution finished in 1 minutes and 30 seconds.

| | MEAN |
|---|---|
| **Average pages/s** | 35.9 |
| **Average requests/s** | 122.6 |
| **Total pages** | 3270 |
| **Total requests** | 11160 |
| **Average Request response time** | 0.1 s |
| **Total request errors** | 0 |
| **Error rate** | 0 |
| **Average Page response time** | 0.264 s |
| **Total throughput** | 20.19 MB |
| **Average throughput** | 1.77 Mb/s |
| **Total users launched** | 10 |
| **Total iterations completed** | 10 |
| **Total action errors** | 0 |
| **Alerts total duration** | 0 % |

### *Scenario 5*

Scenario five is designed to be write heavy with 50% write operations and 50% read. A NeoLoad population was made consisting of 10 users. 5 used the Goal100 user and the other 5 used the View100 user. The non-relational solution took 13 minutes and 56 seconds and the relational solution took 43 minutes and 42 seconds.

| | A | B | % |
|---|---|---|---|
| **Average pages/s** | 1.2 | 11.0 | +817% |
| **Average requests/s** | 2.2 | 18.5 | +741% |
| **Total pages** | 3050 | 9215 | +202% |
| **Total requests** | 5767 | 15450 | +168% |
| **Average Request response time** | 0.799 s | 0.301 s | -62.3% |
| **Total request errors** | 361 | 0 | -100% |
| **Error rate** | 6.3 | 0 | -100% |
| **Average Page response time** | 1.51 s | 0.478 s | -68.3% |
| **Total throughput** | 915.54 MB | 15.8 MB | -98.3% |
| **Average throughput** | 2.79 Mb/s | 0.15 Mb/s | -94.6% |
| **Total users launched** | 10 | 10 | +0% |
| **Total iterations completed** | 10 | 10 | +0% |
| **Total action errors** | 0 | 0 | +0% |
| **Alerts total duration** | 7.3 % | 0 % | -100% |

**RQ2.2: What performance issues are there with the non-database aspects of the system?**
The results in RQ2.1 are not a reflection of the databases. NeoLoad runs its scenarios on the user interface so instead provides results on the performance of the application as a whole. A SQL server and MongoDB comparison is discussed as part of chapter 3. This is a more realistic test as it gives the databases context. The initial motivation for this project was to investigate whether the VRE application would perform better using a non-relational database and the MEAN stack was chose as the alternative implementation. Most previous studies have tested only the database which is an unrealistic approach when it is put in the rehabilitation context. Users are going to be interacting with the database through the application, which can have a significant effect on the performance from the perspective of the end users depending on factors such as design, poor coding, styling and images, etc.

What became apparent quickly with the performance tests in RQ2.1 is that the biggest drain on performance is the unnecessary page requests being sent to the server multiple times and making the total amount of data transferred higher than necessary. Scenario four was reconfigured so that each of the 10 users just updated the same repository item 100 times without clicking back to the repository list each time, it completed in 1 minute and 21 seconds. This significant gain in performance is shown in every statistic:

| | Refresh | No Refresh | % |
|---|---|---|---|
| **Average pages/s** | 5.0 | 39.9 | +698% |
| **Average requests/s** | 19.4 | 136.1 | +602% |
| **Total pages** | 8220 | 3270 | -60.2% |
| **Total requests** | 31950 | 11160 | -65.1% |
| **Average Request response time** | 0.597 s | 0.092 s | -84.6% |
| **Average Page response time** | 2 s | 0.237 s | -88.1% |
| **Total throughput** | 49.85 MB | 4.01 MB | -92% |
| **Average throughput** | 0.24 Mb/s | 0.39 Mb/s | +62.5% |

The non-relational application is structured as follows:

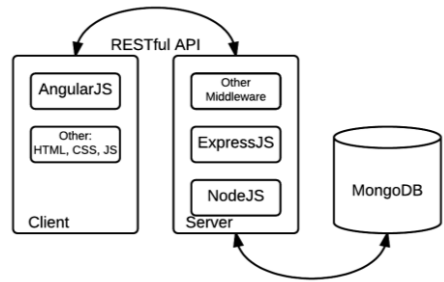

The server request logs and the in-browser developer tools can be used to break the response time down to investigate how each part of the application affects performance. For example, adding a new treatment while using the browser developer tools to record the transaction tells us that the complete operation takes 17.534ms. 14.701ms of which is the actual request and not the browser.

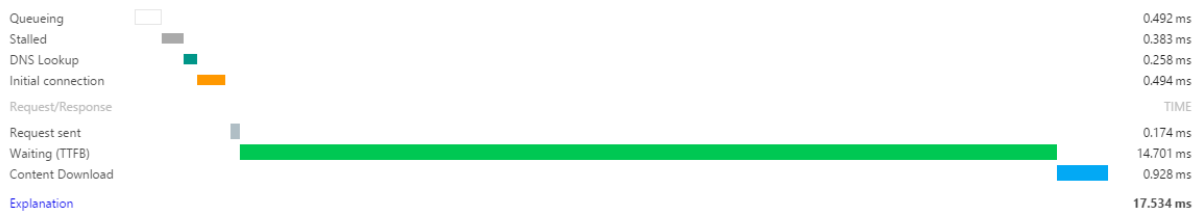

This shows that the longest part of the operation is the Time To First Byte (TTFB). "This time captures the latency of a round trip to the server in addition to the time spent waiting for the server to deliver the response" (Google, 2015). Looking at the server log, the total time for the request to be processed and response to be sent was 9.482ms.

```
POST /api/goal 200 0.818 ms - 207
POST /api/treatment 200 9.482 ms - 211
```

This tells us that 5.219ms of the total time spent was used by the client application. For development, both the client and server code is stored on the same machine. In production, these will each be located on a different server and the request would have to travel over a network which would increase this time. Repeating this test and collating the results gives the following table:

| Operation | Total time (ms) | Client time (ms) | Server time (ms) |
|---|---|---|---|
| Create treatment | 14.722 | 7.383 | 7.339 |
| View treatment | 16.022 | 9.099 | 6.923 |
| Update treatment | 18.451 | 8.896 | 9.555 |
| Delete treatment | 19.776 | 5.288 | 14.488 |
| Upload Repository | 583.012 | 11.641 | 571.371 |
| Assign Content | 15.362 | 6.894 | 8.468 |

All of the operations execute quickly with no clear part of the application that is slower than the other. Each taking a similar time to complete is expected as each of the operations make one call to one service. The size of the object being sent is clearly a factor of overall time as demonstrated by the 'upload repository' operation. It takes significantly longer than the other operations but this is due to it sending and processing an 11MB video to the server, the rest of the operations just send and receive plain text, JSON objects. Even with the large object, the time taken for the client side application to complete is still small. This is because the client just straight away sends the object without doing anything else:

```javascript
$scope.upload = function(files) {
    if (files && files.length && $scope.content) {
        for (var i = 0; i < files.length; i++) {
            var file = files[i];
            Upload.upload({
                url: '/api/repository',
                data: {
                    'name': $scope.content.name,
                    'pat_desc': $scope.content.patient_description,
                    'clin_desc': $scope.content.clinician_description,
                    'category': $scope.catview
                },
                file: file
            }).progress(function (evt) {
                var progressPercentage = parseInt(100.0 * evt.loaded / evt.total);
                $scope.log = 'progress: ' + progressPercentage + '% ' +
                            evt.config.file.name + '\n' + $scope.log;
            }).success(function (data, status, headers, config) {
                $timeout(function() {
                    $scope.log = 'file: ' + config.file.name + ', Response: ' + JSON.stringify(data) + '\n' + $scope.log;
                });
                $scope.success = "Upload successful";
            });
        }
    }
}
```

Currently, both the client and server side applications reside on the same workstation. When put into production, these services will be separated onto their own servers and requests will take longer overall as they will have to travel over a network to and from the server application.

To conclude, although the choice of database impacts on the performance of the application, the database interaction is only a small part of the overall request. Of the current requests being made, around half of the request is spent by the client side application and the other half spent by the server side application and database. This response time will increase again once the client and servers are separated and communicating over a network. Better performance gain will be achieved by increasing the efficiency of these requests and reducing unnecessary requests than working on improving the speed of the database at this time.

## 5.3 RQ3: What is the relative impact of relational and document database design on overall system scalability?

John Engates, the CTO of Rackspace defines scalability as a desirable property of a system which indicated its ability to either handle growing amounts of work in a graceful manner, or to be readily enlarged as demands increase (Engates, 2008). Web applications today should be designed for scalability from the outset, especially when as is the case with the VRE, the application aims to cater to a large population. Applications may be scaled either vertically or horizontally. Scaling vertically is the easiest method as it just involves adding more resource to the current hardware (e.g. more memory, disk space). However, hardware can only be scaled vertically for as much as the motherboards allow and can quickly become expensive. Horizontal scaling or scaling 'out' is where the database is run over multiple servers. The initial application operated on top of a RDBMS, which have a reputation for being difficult to scale out. This is a common misconception.

**Typical solutions for RDBMSs'**
Here are the four most common ways to scale a Microsoft SQL Server (Microsoft, 2012):

1. Scalable shared databases – a storage area network (SAN) holds the data for up to eight SQL server instances on different servers. SQL servers maintain the database locks in memory meaning that this solution cannot work with locks. This solution is great for read only applications such as data warehouses or reporting databases.
2. Peer to peer replication – each server has a copy of the database. When a write operation is executed on a server, the change is propagated out to the rest. This solution does not support conflict resolution and therefore should only be used where only one server is updating a set of records.
3. Linked servers with distributed queries – the application is split by functional area and each area has its own database. Synonyms are then granted so that other servers can query the data. However, querying remotely is significantly more expensive and should only be done when absolutely necessary
4. Data dependant routing – the data is split so that for example, the first five thousand patients use server one, the next five thousand use server two etc.

For the VRE application, scaling the existing relational application would not be suited to just one of these above methods and would require a combination. Clinicians manage many patients within their geographical district. Clinicians are not likely to make changes to any patients outside of their district so therefore the peer to peer replication method seems like a good choice. However, this would mean that there are multiple copies of the repository that need to be maintained. The existing application stores content on the server and cannot be integrated with a content discovery network (CDN) so therefore the large repository, central to the application would be replicated over each server wasting a lot of storage space. The repository would be more suited to the scalable shared database solution where updates are made occasionally and most of the traffic originates from read operations. The content would all be stored in the central SAN. However, this solution isn't ideal for the rest of the application that would require constant read and write transactions to be taking place.

**Scaling the new application**
Instead for the new application, a non-relational database was chosen. Non-relational databases sacrifice some of the ACID (Atomic, Consistent, Isolated, and Durable) properties needed for relational databases (Hurst, 2010) in return for a more scalable solution, instead replacing ACID for BASE (Basic Availability, Soft-state and Eventually consistent). The most common way of scaling mongo is by using sharding. Sharding works in a similar way to the data dependant routing solution for SQL servers. A database is split in to multiple shards, which most commonly are multiple physical servers set up as a replica sets (MongoDB [II], 2015). The data is split amongst these shards and the application no longer interacts with the database but instead a router is used called mongos.

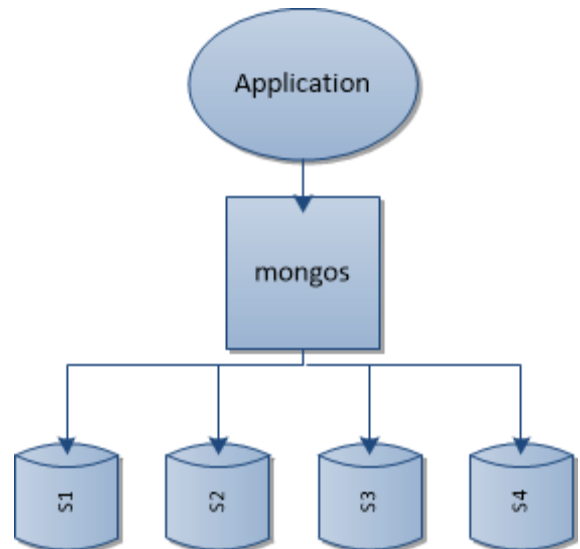

*Application architecture using sharding*

Data is split over the shards using a defined range based shard key. This means that if the patient ID is used as the shard key, the first five thousand patients may be stored on shard one, the next five thousand on shard two etc. When querying the data, if the patient ID is included in the query then the router knows to only search the shard that ID is registered too, providing the user with a quick response. If the query does not include the patient ID then the router searches each of the shards to answer the query (MongoDB [II], 2015). Once a shard key has been decided upon, all records must include the shard key for the record to be legal. Otherwise the router does not know which shard to place the data into. This means that the patient's telephone number would not make a suitable shard key as not all patients may have a telephone. Multiple routers can be set up and if one fails then the application just uses a different one. If set up correctly, Mongo is excellent for scaling as it just requires extra shards being added to the cluster. However, Mongo has been criticised for the way it achieves its impressive benchmarks. When a write command is executed on Mongo, it is not written anywhere. Instead it is staged to be written at a later time. If a problem occurs before the data is permanently written, it is lost forever.

The database is not the only important aspect to consider when designing an application for scalability. In the 2008 LinuxWorld conference, John Engates presented the seven stages of scalability, describing how scaling becomes increasingly painful until the database is partitioned meaning that some features get their own database (Engates, 2008). Netflix (Mauro, 2015) and Twitter (Schuller, 2014) follow this similar architecture, splitting their applications up in to multiple microservices. Adrian Cockcroft defines the microservices architecture as a service-oriented architecture composed of loosely coupled elements that have bounded contexts (Mauro, 2015). This means that each service can scale independently of the others and leads to higher performance overall as resource can be allocated appropriately where needed. The new application utilises this multiservice architecture, preparing the application for future growth. It also is designed to be easily integrated with a CDN so that the repository content can also scale independently of the application.

**Impact of scaling solution chosen**

The main impact of choosing a non-relational solution over a relational solution for the VRE is that once initially set up, the non-relational solution scales as needed without requiring any changes to the application code. This can save the organisation a significant amount of money in development costs on top of the costs already spent on the new hardware. The results of section 5.1 also found that the non-relational solution uses a lot less processing power than the relational solution. This means that the application is less likely to need to be scaled out as soon as the relational solution would need to be. Again, this means delaying the need to scale out and saving costs which are crucial to a new organisation with little capital. The CPU heavy, relational solution will outgrow its starting hardware quicker as the number of users increases. Once it needs to scale out, the process is likely to require application changes. This is especially true if multiple databases are used to satisfy the different functional areas of the application as mentioned previously.

The research here has led to the proposal that the best way to scale the VRE application initially would be using the microservices architecture mentioned above. The database, server application and client application can all be separated out into independently managed services that can each scale when necessary and independently of the others. Also, a third party, hosted CDN can be used to provide the repository content and the client applications JavaScript files, stylesheets and images. This gives the VRE the infrastructure needed to provide users with instant high performance as a subscription service negating the need for a large initial investment to set up and maintain a CDN themselves.

# 6. Discussion, evaluation and conclusion

## 6.1 Discussion

The investigation set out to answer three questions; what are the technical constraints and particular requirements of the problem? What effect on performance does choosing a non-relational solution over a relational solution provide? And what impact does choosing a non-relational solution over a relational solution have on scalability?

To answer the first question, the materials from the first project had to be read and reviewed. This included the notes from initial meetings and design documents. Other than the functional requirements for the system stated in appendix A, there were four main functional requirements:
- The look and feel had to be natural and fit in with the clinicians busy schedule
- The application had to comply with the NHS information governance (NHS, 2015)
- The application needed to be able to handle a large amount of traffic
- The application needed to be able to scale out to handle this traffic

The first project only addressed the first two of these requirements which sparked the initial motivation for this second project. The answer to the first question describes other features that have been implemented to contribute to the performance and scalability of the new application, such as easy integration with CDN's and functional modularisation.

The second question is broken into two parts, the first looks at application performance and the second is a review of the performance of the non-database aspects of the application. The tests were chosen, inspired by the Yahoo Cloud Services Benchmark (Cooper et. al, 2010) and reveal some interesting issues. At a glance it is clear to see that the non-relational solution performs better overall but flaws in design affect the application significantly. In the first test, the non-relational application took 51 minutes and 17 seconds to insert 1000 clinicians into the database. The relational solution took only 2 minutes and 43 seconds. The problem here was that the non-relational solution refreshed the page after every insert which led to the browser re-requesting all the pages needed for the application each time. Most of these requests would return a 304 'Not Modified' response but the request would still take time. As it is a single page application, the pages are supposed to be requested once during login and then remain on the client side for when they are needed. This led to the non-relational solution making 82,980 requests whereas the relational application only made 3110. This design flaw was because the functionality for creating accounts was developed at the start of the project when the technology was new to the researcher. For this reason, the second test was similar in that it adds 1000 records but was done on the clinician's interface which was developed further into the project.

Test two shows a significant performance advantage over the relational solution. The test completed in only 32 seconds where the relational solution took 8 minutes and 32 seconds. Also, without the page refresh the total number of requests was 3,070, less than the relational solutions 4956. As the application is now only sending the requests to insert the goals, the total throughput is significantly reduced to only 1.07MB (it being 99.89MB on the first test). This is what makes it so fast to process, the design of the operation means that the least amount of data possible is sent to the server and therefore the test takes a lot less time to complete. Also, this test showed the first indication that the relational solution was CPU heavy. The test on the relational application had an average CPU load that was double the average of the non-relational application. The average memory allocation between the two solutions was similar but the non-relational solution was more consistent throughout the experiment. This heavy CPU load on a relatively low load of data indicates that it is likely that the relational solution would need to be scaled out quicker than the non-relational solution.

| % | SQL MIN | MEAN MIN | SQL AVG | MEAN AVG | SQL MAX | MEAN MAX |
|---|---|---|---|---|---|---|
| Controller/CPU | 7 | 6 | 80.1 | 42.4 | 100 | 87 |
| Controller/Memory | 43 | 54 | 54.5 | 58.1 | 70 | 59 |

The third test was designed to simulate everyday usage of the application with 90% of the traffic being read operations and 10% write. The non-relational solution took 22 minutes and 56 seconds and the relational solution took almost double this at 43 minutes and 41 seconds. Interestingly in this test, the non-relational application made a lot more requests to its server, 25,354 in total with the relational solution only making 6,860. However, again it is only sending minimal data with most of the responses being 304 'Not Modified' responses. This 97.8% reduction in data being sent is why the non-relational application is able to perform faster.

Test four updated the names of 1000 repository items over ten users. The non-relational solution finished in 1 minute and 30 seconds, even with the highest total throughput of tests 2, 3, 4 and 5. The relational solution quickly overloaded the CPU and the workstation itself shut down so there are no results to compare. However, it does again indicate that the relational solution would quickly need to be scaled in a production environment.

Finally, test 5 simulated a write heavy load on the application to test how the application would perform under a large amount of write operations. The non-relational solution took 13 minutes and 56 seconds and the relational solution took 43 minutes and 42 seconds. Again, as with the previous tests, the amount of requests for the non-relational solution was significantly higher, but the test finished faster as the total throughput was 98.3% less than the relational application. 361 errors occurred on the relational solution but had minimum effect on performance. However, again the CPU overloaded.

The second part of the second research question looks at what factors of the application affect performance. Currently, the database is the fastest part of the application with the majority or request times being nearly evenly split between the client application and the server side. However, both the client and the server side code currently reside on the same server. Once in production and having to travel over a network to communicate, the large amount of requests that the non-relational application sends will pose a large threat to the performance of the application. For this reason, the investigation of RQ2.2 identified that more work to reduce these requests will be needed on the non-relational application before it is ready for production.

The third research question looks at the impact different technologies have on overall scalability. The resource was not available during the project to test different scalability techniques so instead a review of material was undertaken and a proposal made based on that review. The review looked at the most common methods for scaling out a SQL database and found not one method satisfied all of the uses that are required for the VRE application. However, MongoDB users manage their shards through a graphical interface that allows you to simply add new servers as they are needed. The pattern used by Mongo is similar to the data dependant routing method that relational applications use but when used in conjunction with a CDN for the repository, it is well suited to the VRE. Overall, the investigation for RQ3 found that the VRE application is best suited to a microservices architecture made up of small, independently scalable pieces such as the server side application, client side application, CDN and the database.

## 6.2 Evaluation

Overall, the results confirm initial suspicions that the application could perform extremely well as a non-relational solution. The flexible data structure it provides would mean that additions and modifications can be easily made as the application grows without the developer needing intricate knowledge of how the rest of the application is designed. A non-relational solution combined with other design features such as a CDN for the repository also gives the application the capability of scaling easily and cost effectively. However, with all the advantages provided by a non-relational technology, a good understanding of the disadvantages is crucial when developing non-relational applications. Fortunately, having a system that complies with BASE over ACID is not too important in the VRE application as it is not a critical system and so does not matter if the data is not consistent. A good understanding is needed of the particular technology as well; Mongo achieves its impressive benchmarks by not writing to the database directly. Instead, changes are staged to be written at a later date. If there is problem writing, the user making the change may have had confirmation that the operation completed successfully (MongoDB [III], 2015). An awareness of issues such of these should be had by all developers as considerations need to be made throughout the application.

The main point learnt during this project is that the technology chosen is not nearly as important as the design decisions made when designing for performance. In both applications, the database interaction was the fastest part of all operations. What slowed the operations down was the amount of data transferred. Single page applications download the majority of scripts and stylesheets when the user first logged on. A major problem with the non-relational VRE application is that some functionality forces a page refresh which downloads all these files again which slows performance. Designing the application to not do this would achieve greater performance results than adding a faster database. The project concluded that the best solution for the VRE application would be to design it in a series of small and independently scalable parts where each part can be also tuned for performance independently. This allows each part to scale as is needed and performance can be compartmentalised and assessed, unlike the relational application that is wrapped up into a compiled library that we cannot see the inner workings of.

## 6.3 Further research

To follow on from this project, the testing can be broken up further to see what differences in performance there are between just the two databases without an application on top. Most of the material analysed has performed the Yahoo Cloud Services Benchmark test against multiple non-relational databases to find out which is the quickest. There is little academic work comparing relational with non-relational databases.

Testing on a workstation with a higher specification is desired as the CPU used in this investigation could often not handle the CPU requirements of the relational application.

Also, the applications could both be scaled out to multiple servers and tested. For the non-relational application, the database can be scaled into a Mongo cluster, then the client and server applications could be separated onto different servers. A CDN could also be used to store all of the repository content and client applications files. If this project were to be carried out again, a CDN would have been used for the client content from the start. There is material online that demonstrates how to set up Google App Engine as a private CDN (Krohn, 2008).

## 6.4 Project management

From the outset, it was understood that this project would be development heavy and so the majority of time was allocated to the development of the application. Below is the initial plan for the project as submitted in the mid-term report:

| Task | Target Completion |
|------|-------------------|
| Proposal hand in | 11/06/15 |
| Development start | 11/06/15 |
| Mid-project report | 23/07/15 |
| Development end | 12/08/15 |
| Test environment setup | 16/08/15 |
| Performance testing | 23/08/15 |
| Report hand in | 17/09/15 |
| Demonstration | TBC |

It shows that the development was planned to finish on the 12th August, then four days were to be spent setting up the test environments, a week testing and four weeks of writing the report. These target dates were kept to, with development being cut off on the 12th with still one planned feature outstanding (see appendix A). However, leaving only one week for testing was naïve in hindsight. The testing took longer and was something that was continually carried out as the report was written. This has led to there being a large application but only a small subset of functionality tested. The five scenarios tested in section 5.1 cover a broad amount of operations and are sufficient for the project but it means that the features that have been developed and then left feel insignificant. A more appropriate use of time would have been to spend a small amount of the project developing a handful of features and then the rest of the project testing and making changes to make them more efficient. Initially, the project planned to deliver all 27 features of the application but this was cut down to 15 early on in the project as unfamiliarity with the technology and other commitments throughout the project meant that progress was slower than anticipated. Despite this, the project maintained organised throughout. As the development was a copy of an existing application, the requirements were well defined from the outset. GitHub (GitHub, 2015) was used as the code version manager and all documentation was managed through Google Drive (Google [II], 2015). Links to the code and documentation can be found in appendix B.

## 6.5 Conclusion

To conclude, the project set out to discover if the VRE application would be better suited to an application built on top of a non-relational database. Overall, it did achieve this goal. A solution was designed and built to be tested. During the process, other design choices came to light and they have amalgamated into a proposal for how the application should be designed going forward. The performance tests carried out show that the new application performs quicker and more efficiently although more work needs carrying out to further reduce unnecessary requests and increase performance more. The technology used for the new application is more flexible and easier to scale. The trade-offs required of using a non-relational database are identified and discussed. Concluding that as the VRE does not handle critical data, these trade-offs are acceptable given the benefits to the application that accompany them. However, although all of the testing carried out so far suggests that the non-relational application is a better solution, more extensive testing needs to take place on a larger set of the applications functionality to ensure there are not any hidden design flaws such as the one identified in scenario one in section 5.1. The application would also benefit from a prototype being created for performance testing using the distributed proposal suggested in section 5.3.

## *Appendix*

### Appendix A. Feature Catalogue

Here is a full feature catalogue for the core application. Time constraints meant that only key features were developed during the project. Included is a comparison of which features have been developed as part of each application.

Key:

    ✔ - Included

    ✘ - In development/Intended Deliveries

    ✘ - Not included

| Feature | .NET App | MEAN App |
|---|---|---|
| **Common Functionality** | | |
| Log In | ✔ | ✔ |
| Log Out | ✔ | ✔ |
| Change Password | ✔ | ✘ |
| View Profile | ✘ | ✘ |
| Edit Profile | ✘ | ✘ |
| | | |
| **Administrator Interface** | | |
| CRUD Account | ✔ | ✔ |
| Assign Patient to Clinician | ✔ | ✔ |
| CRUD Category | ✔ | ✔ |
| CRUD Content | ✔ | ✔ |
| | | |
| **Clinician Interface** | | |
| View Patient List | ✔ | ✔ |
| View Patient | ✔ | ✔ |
| CRUD Goals | ✔ | ✔ |
| CRUD Treatments | ✔ | ✔ |
| CRUD Information | ✘ | ✔ |
| Lock Patient Account | ✔ | ✘ |
| Assign Self To Patient | ✔ | ✔ |
| CRUD Goal Comments | ✔ | ✘ |
| Browse Repository | ✔ | ✔ |
| CRUD Private Repository | ✔ | ✘ |
| Assign Content to Patient | ✔ | ✔ |
| | | |
| **Patient Interface** | | |
| View Clinicians | ✔ | ✘ |
| View Goals | ✓ | ✗ |
| View Treatments | ✓ | ✗ |
| View Information | ✗ | ✗ |
| CRUD Comments | ✗ | ✗ |
| | | |
| **Extra Features** | | |
| High security accounts | ✗ | ✗ |
| Reporting functionality | ✓ | ✗ |

**Appendix B. Project Resources**

**Source code**
The source code for this project is stored on GitHub: https://github.com/zedrem/VRE

**Initial Application**
Source code: https://github.com/zedrem/VRE_Relational

Final Report:
https://drive.google.com/file/d/0B18dn1gn9SHATDFTU0hPdnA4d00/view?usp=sharing

Design Documents:

| Overview Specification | https://docs.google.com/document/d/1ajloVmxy704TdNRBcTccPktwmD1kED_99oZzxkmD0Uo/edit?usp=sharing |
|---|---|
| Login Functionality | https://docs.google.com/document/d/1X4mXDEx-Q56OL2nw2LvhMPebMxAb08FMeqMtGQjq70Q/edit?usp=sharing |
| Central Repository | https://docs.google.com/document/d/1AMb1HFU58WNDrmDGWEfq552HElzV7c1EeVJgJL5caZQ/edit?usp=sharing |
| Patient Management | https://docs.google.com/document/d/1qZwaUyMFCbT_8gQDDrg7wNiO_ADHSxSRatjJzaUtE7M/edit?usp=sharing |
| Assigning Content | https://docs.google.com/document/d/1-JiFm-0NyM_ctD-nfs2qQcw8l_Pp3NbfZym4qeOXeTo/edit?usp=sharing |

**Appendix C. Complete walk through for uploading content to the repository**
To fully demonstrate how the components of the MEAN stack work together, here is a walkthrough of how an administrator uploads content to the central repository:

1. User inputs the contents' information into the 'Create Content' modal and selects the file to be uploaded.

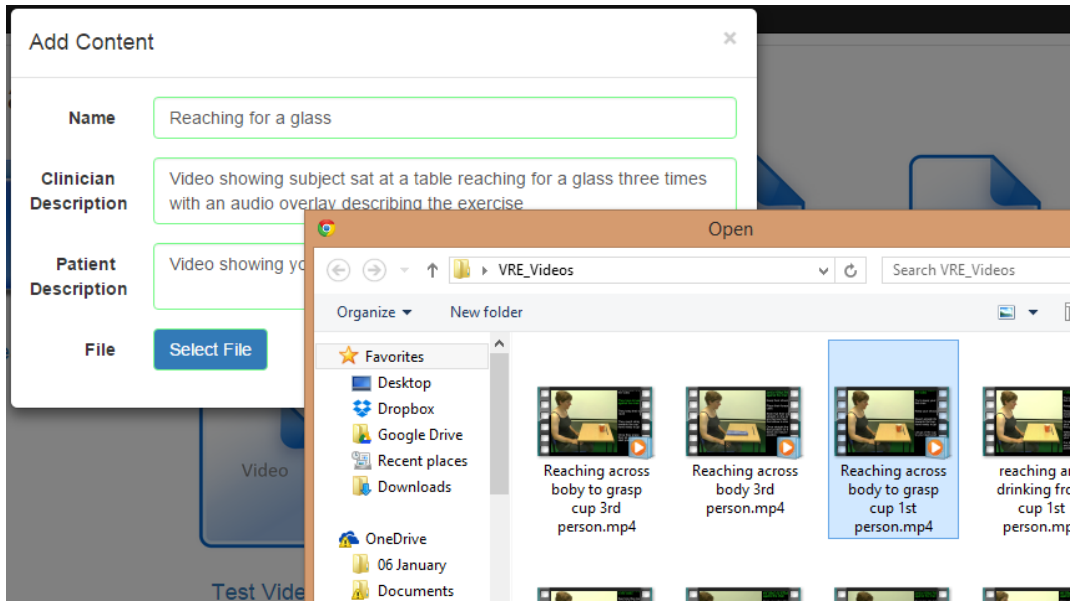

2. The content upload is different to the rest of the application in that it does not use the ngResource service to interact with the repository. Instead it interacts directly using ngFileUpload (Farid, 2014), an open source upload tool for uploading files in Angular.

```
$scope.upload = function(files) {
    if (files && files.length && $scope.content) {
        for (var i = 0; i < files.length; i++) {
            var file = files[i];
            Upload.upload({
                url: '/api/repository',
                data: {
                    'name': $scope.content.name,
                    'pat_desc': $scope.content.patient_description,
                    'clin_desc': $scope.content.clinician_description,
                    'category': $scope.catView
                },
                file: file
            }).progress(function (evt) {
                var progressPercentage = parseInt(100.0 * evt.loaded / evt.total);
                $scope.log = 'progress: ' + progressPercentage + '% ' +
                    evt.config.file.name + '\n' + $scope.log;
            }).success(function (data, status, headers, config) {
                $timeout(function() {
                    $scope.log = 'file: ' + config.file.name + ', Response: ' + JSON.stringify(data) + '\n' + $scope.log;
                });
                $scope.success = "Upload successful";
            });
        }
    }
}
```

3. Hits the API and the route called the upload function in the repository controller

```
module.exports = function(app){
    app.route('/api/repository')
        .get(repository.list)
        .post(accounts.requiresLogin, repository.upload);

    app.route('/api/repository/:repositoryId')
```

4. Moves the file from the temporary store used by ngFileUpload to its permanent store in the repository, creates a Content object using the Content model and then persists the data sent and meta data to the Content collection in the database using the save() function provided by Mongoose (Mongoose, 2015).

```
exports.upload = function(req, res, next){
    var file = req.files.file;
    var content = JSON.parse(req.body.data);
    var tmpPath = file.path;

    //move file to repository
    var path = tmpPath.split('\\');
    var uniqueFileName = path[path.length - 1];

    console.log(uniqueFileName);
    var newPath = config.VRE_GLOBAL_REPOSITORY + "\\" + uniqueFileName
    console.log(newPath);
    fs.move(tmpPath, newPath, function(err){
        if(err){
            console.log(err);
        }else{
            //console.log("success");
            var con = new Content();
            con.name = content.name;
            con.type = file.type;
            con.patient_description = content.pat_desc;
            con.clinician_description = content.clin_desc;
            con.category = content.category;
            con.path = newPath;
            con.creator = req.user;

            con.save(function(err){
                if(err){
                    return next(err);
                }else{
                    res.json(con);
                }
            });
        }
    });
};
```

5. Success message displayed.

| | |
|---|---|
| Upload successful | |
| **Name** | Reaching for a glass |
| **Clinician** | Video showing subject sat at a table reaching for a glass three times |

**Appendix D. Performance test results**

Below are links to test reports generated by NeoLoad for each of the five scenarios in section 5.1:

Scenario one:
https://drive.google.com/file/d/0B18dn1gn9SHAcWM1Wm1nTUZ5dGc/view?usp=sharing

Scenario two:
https://drive.google.com/file/d/0B18dn1gn9SHAandiOUtEemFUemM/view?usp=sharing

Scenario three:
https://drive.google.com/file/d/0B18dn1gn9SHAY1JyS3hJczIxNTg/view?usp=sharing

Scenario four:
https://drive.google.com/file/d/0B18dn1gn9SHAc1ZEQkZDSEUzNG8/view?usp=sharing

Scenario five:
https://drive.google.com/file/d/0B18dn1gn9SHAZlV6V0JFcHVQSEk/view?usp=sharing